\documentclass{optica-article}

\journal{opticajournal} 

\articletype{Research Article}
\usepackage{graphicx}   
\usepackage{lineno}

\usepackage{cite}
\usepackage{amsmath}
\usepackage{graphicx}
\usepackage{float}
\usepackage{placeins}
\usepackage{wrapfig}

\graphicspath{ {./images/} }
\usepackage{lipsum}     

\begin{document}

\title{Subwavelength micromachined vapor-cell based Rydberg sensing}

\author{Avital Giat,\authormark{1} Kfir Levi,\authormark{1} ,Ori Nefesh\authormark{1} and Liron Stern\authormark{1,*}}

\address{\authormark{1}The Faculty of Science, The Center for Nanoscience and Nanotechnology, Institute of Applied Physics, The Hebrew University of Jerusalem, Jerusalem 91904, Israel \\

}

\email{\authormark{*}liron.stern@mail.huji.ac.il} 


\begin{abstract*} 
In recent years, micromachined vapor cells have been revolutionizing the field of chip-scale quantum sensors such as magnetometers and atomic clocks. In parallel, Rydberg atomic quantum sensing has emerged as a powerful technique for broadband, non-invasive and ultra-sensitive electrometry. Yet, to date, Rydberg sensing has largely been limited to glass-blown, cm-scale vapor cells. Here, we perform Rydberg spectroscopy using a wafer-scale fabricated Pyrex–Si–Pyrex cell with mm-scale dimensions. The Rydberg spectroscopic line is characterized with respect to critical parameters such as temperature, the frequency and amplitude of the applied radiofrequency field, light intensity, and the spatial position of the interrogating beam. Our study reveals lineshapes directly influenced by a complex landscape of electrostatic fields with values up to approximately 0.6 V/cm. By controlling key parameters, we were able to reduce the effect of these internal electric fields, and demonstrate the detection of RF fields with a sensitivity as low as 10 \(\mu\)V/cm. These results highlight the potential of micromachined vapor cells for sub-wavelength electromagnetic field measurements, with applications in communications, near-field RF imaging, and chip-scale quantum technologies.
\end{abstract*}

\section{Introduction}
Rydberg atoms are a class of atoms in which one or more valence electrons are excited to exceptionally high energy levels, causing many of their physical properties to be greatly amplified~\cite{gallagher1988rydberg, edelstein1979rydberg, haroche1985radiative}. These atoms exhibit notable features such as exceptionally high polarizability, long lifetimes, large atomic radii, huge dipole moments, and closely spaced energy levels. In addition, Rydberg atoms exhibit large transition dipole moments between neighboring levels, enabling strong coupling of electric fields to frequency differences spanning the microwave (MW), radio frequency (RF), and terahertz (THz) ranges. ~\cite{urban2009observation, barredo2014demonstration}.  These distinctive properties position Rydberg atoms as a powerful platform  enabling novel applications in quantum technology, particularly, rendering them an appealing choice as electric-field quantum sensors supporting a broad range of frequencies.

Indeed, high‐quality, sensitive RF field sensing has been achieved using Rydberg atomic vapor cells, leveraging the ability to split the Rydberg lineshape—often referred to as Autler–Townes (AT) splitting ~\cite{sedlacek2012microwave, liu2023electric}. Rydberg sensors offer significant advantages over traditional methods (i.e., classical antennas) in many respects. First, in terms of detection sensitivity, Rydberg atoms can potentially surpass the thermal limit of conventional antenna technology. Recent years have witnessed significant efforts toward this goal, with demonstrations of Rydberg sensing exhibiting amplitude sensitivity floors as low as\( \sim {nV}/{cm} \) ~\cite{Jing2020}.Additionally, Rydberg sensors offer all-optical detection and, due to their relatively low content of RF-scattering materials, hold promise for non-invasive measurements. Other unique properties of Rydberg sensors include the ability to be adaptively tuned to measure desired frequencies across a vast energy landscape, and the possibility of field imaging with sub‐wavelength resolution due to the size and composition of the cells. These advancements have enabled a wide range of demonstrations, including communication ~\cite{adams2019rydberg, gong2024rydberg, meyer2018digital}, RF phase detection ~\cite{anderson2020rydberg, simons2019embedding}, RF imaging and mapping ~\cite{simons2019embedding, holloway2014sub, adams2019rydberg}, and angle‐of‐arrival estimation ~\cite{robinson2021determining, richardson2025study, bottomley2024sub}.

When considering vapor‐cell technologies, recent years have witnessed the emergence of micromachined, wafer‐scale approaches for fabricating vapor cells that promise unprecedented miniaturization as well as exceptional compatibility with photonic components. ~\cite{schmidt2005electromagnetically, goban2014atom,Kitching_2016 }. These micromachined vapor cells are fabricated using silicon‐based processes, in contrast to conventional centimeter‐scale cells produced via traditional glass‐blowing techniques. This advanced approach, which leverages semiconductor manufacturing methods, enables the creation of compact, cost-effective, planar vapor cells with high repeatability, mass producibility, and efficient photonic integration. While the earliest work in micromachined cells focused on realizing chip‐scale atomic clocks ~\cite{knappe2005chip, kitching2018chip}, recent progress has broadened their applications to include chip‐scale atomic magnetometry ~\cite{levi2023remote, kitching2018chip}, optical atomic clocks~\cite{newman2019architecture, martinez2023chip}, and frequency microcomb spectroscopy ~\cite{stern2020direct, scalari2019chip}. Even more recently, micromachined vapor cell technology has facilitated integrated waveguide–atom interactions~\cite{stern2013nanoscale, zektzer2021nanoscale}, representing a key advancement toward scalable photonic–atomic architectures~\cite{PICmicroMachined}. Yet, for the most part, Rydberg atomic sensors have relied on glass‐blown cell technology and centimeter‐scale apparatuses, thereby limiting their utility and potential for noninvasive, subwavelength‐resolution electrometry.
 
Here, we present a chip-scale electric-field sensor based on a micro-machined vapor cell for all-optical, sub-wavelength RF sensing (see Figure~\ref{fig:fig1}). Through careful examination and systematic characterization of key parameters, we successfully achieve narrow-linewidth Rydberg spectroscopy in these ultra-compact vapor cells. The resulting spectroscopic response, depicted in Figure~\ref{fig:fig1}(c), enables high-sensitivity RF detection with estimated sensitivities  as low as \( 10\ \mu V/\text{cm}\). We investigate how temperature, laser intensity, RF frequency, and beam position affect the Rydberg lineshape and sensitivity for RF sensing, as well as identify a spectroscopic signature consistent with a constant surface-induced electric field within the micromachined vapor cell. We also examine the origin of these electrostatic fields and discuss strategies to mitigate their impact on electrometry.  Our findings contribute to a deeper understanding of Rydberg physics and the electrostatic fields in micromachined cells, with a particular significance for recently  emerging two-photon-based chip-scale optical atomic clocks  ~\cite{kitching2024next, newman2019architecture}, and advance the development of chip-scale based RF sensing technologies by elucidating the intricate interactions between Rydberg atoms and local electric fields in micro-machined vapor cells.

\section{Concept of chip-scale Rydberg sensing}

\begin{figure}[ht]
\includegraphics [width=1\linewidth]{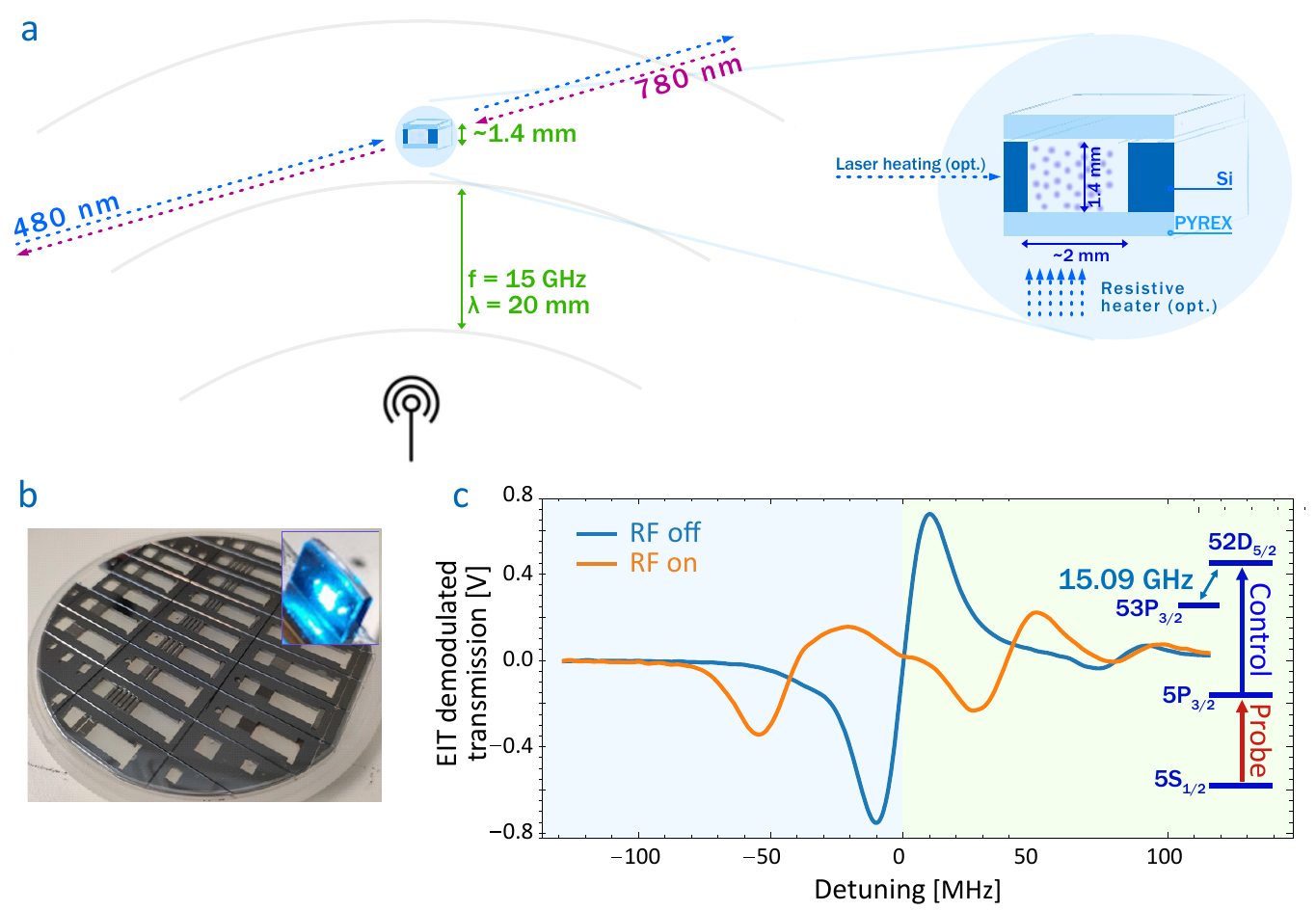}
\caption{\label{fig:fig1} \textbf{Chip-scale Rydberg Quantum Sensing} (a) Conceptual artistic depiction of Rydberg electrometry utilizing a mm-scale micromachined vapor cell.
(b) A fabricated multi-stack bonded wafer (Pyrex-Si-Pyrex) which allows the realization of multiple stand-alone mm-scale vapor cells. (c) Electromagnetically induced transparency (EIT) Rydberg spectroscopy, featuring a demodulated transition line in the form of a gaussian derivative in a micromachined cell. When an RF field is applied, the Rydberg level is coupled to its neighbor, resulting in Autler-Townes (AT) splitting.
 }
\end{figure}

Figure\ref{fig:fig1} illustrates the core concept of our chip-scale subwavelength Rydberg sensing apparatus. As shown in Figure \ref{fig:fig1}(a), we employ a two-photon excitation scheme using two counter-propagating lasers to excite and probe Rydberg levels in a micromachined vapor cell, featuring a cavity volume of \(2 \times 2 \times 1.4\, \  \text{mm}^3\) connected to a small round window holding a dispenser pill. In experiments reported here, we retained a frame of approximately  \( 11  \ \text{mm} \ \times \ 7  \ \text{mm}\), however, we routinely fabricate similar  cells with frames nearly as small as the interaction cavity, in which we have also observed clear Rydberg spectroscopic signals. We note that the interaction cavity dimensions are more than an order of magnitude smaller than the approximately 20 mm wavelength of the applied RF field. Our cells are diced from a custom-fabricated wafer,  which photograph is presented  in Figure ~\ref{fig:fig1}(b).  The wafer itself is composed of an anodically bonded multi-structure composed from Pyrex–silicon–Pyrex layers. As detailed below, by targeting a state with a principal number \textit{n} of 52, we are able to perform pump-probe spectroscopy with lineshapes approximately 20 MHz wide (see Figure \ref{fig:fig1}(c)). The measured demodulated signal takes the form of a Gaussian-derivative lineshape that splits under the RF field, revealing the characteristic AT splitting. Our ultra-compact footprint, supported by a mature and cost-effective platform, enables spatial resolution below \(<\lambda/10\) spatial resolution. We further present a detailed spectroscopic analysis that reveals the lineshape, and sensitivity of the Rydberg response to key system parameters.

We achieve Rydberg spectroscopy by implementing ladder-type Electromagnetically Induced Transparency (EIT), as illustrated in Figure\ref{fig:fig1}(c). Two laser sources at 780\,nm (red) and 480\,nm (blue) serve as the probe and pump lasers, respectively. The probe laser is tuned to coincide with the 5S-5P Doppler broadened D2 lines of \(^{85}\mathrm{Rb}\). Meanwhile, a counter-propagating pump laser scans around the transition between the 5P state to the 52D Rydberg level. When both lasers satisfy the two-photon resonance condition, a narrow transparency window emerges in the probe absorption spectrum. When an RF field couples the \(52D _{5/2}\)  Rydberg state to the \(53P_{3/2}\) state, the RF field dresses the Rydberg state which is imprinted upon the probe signal. This results in a characteristic AT-doublet in the probe spectrum, with the separation between peaks given by:

\begin{equation}\label{eq:ATsplitting}
\Delta f = \Omega_{RF} = \frac{d_{RF}}{\hbar} \bigl| E_{RF} \bigr|,
\end{equation}
where $\Omega_{RF}$ is the Rabi frequency of the RF transition, $d_{RF}$ is the dipole matrix element, and $\bigl| E_{RF} \bigr|$ is the amplitude of the RF field. Since  $d_{RF}$ is accurately known, 
measuring \(\Delta f\) (see Figure~\ref{fig:fig1}) provides a direct method to quantify the local RF field-amplitude. In addition to RF sensing, Rydberg atoms are highly sensitive to DC electric fields due to their large polarizability , which scales as (\(\alpha_0 \propto n^7\)), which leads to a prominent DC Stark shift~\cite{haroche1985radiative,ma2022measurement, duspayev2024high}. Consequently, the same Rydberg atomic vapor cell employed as a broadband RF sensor can likewise perform high-quality measurements of DC electric fields ~\cite{PhysRevResearch.6.023138}.  In the presence of a DC field the transition resonance frequency shifts according to what is known as the DC stark shift:
\begin{equation}\label{eq:DCsplitting}
\Delta f = \frac{1}{2} \,\alpha_0 \, \mathrm{E_{DC}}^2,
\end{equation}
where $\alpha_0$ is the polarizability of the Rydberg state, and $\mathrm E_{DC}$ is the external DC field strength.

Due to the atoms’ proximity to the micromachined cell walls, the sublevel manifold must be considered, as its degeneracy is lifted by electric fields originating from surface charges. Such sublevels, are specified by the magnetic quantum number \( |m_J| \). In the absence of an external field perturbation these sub-levels are degenerate. However, when an external field is introduced, this degeneracy is lifted, and the sub-levels experience distinct energy shifts. This occurs because each sub-level interacts differently with the applied field, primarily due to variations in their polarizability, \( \alpha_0 \), and their dipole moment. The occurrence of non-uniform shifts among these sublevels will cause the overall lineshape to shift, broaden, and, in strong fields, even split~\cite{li2023super}. 

\begin{figure}[h]
\includegraphics[width=1\linewidth]{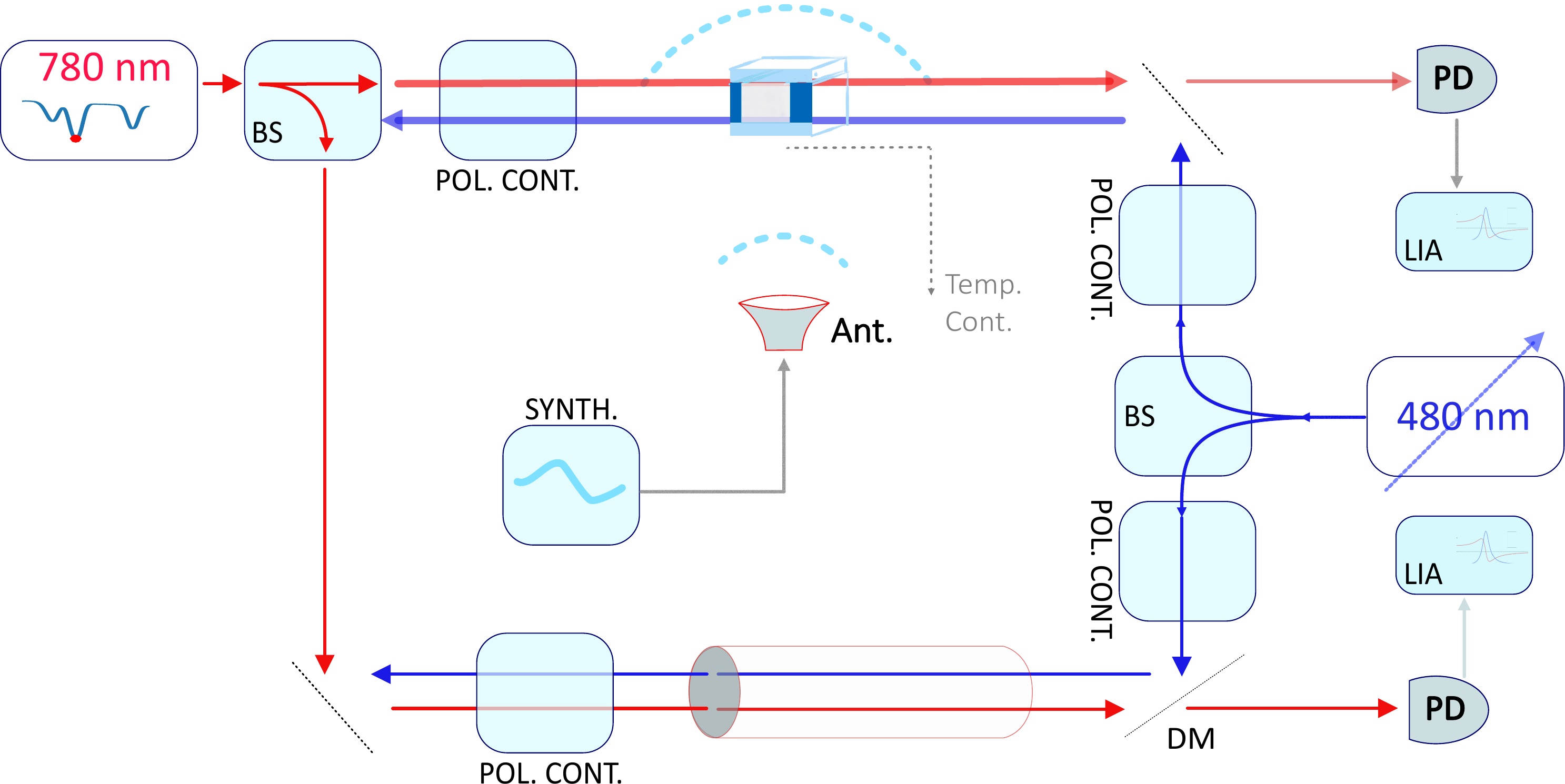}
\caption{\label{fig:setup} \textbf{Schematic illustration  of experimental system.}  Two tunable lasers at 780 nm and 480 nm  serve as the probe (illustrated in red) and pump (illustrated in blue). The lasers are split into two parallel beam paths, where the bottom path produces the reference signal using a glass-blown cm-scale vapor cell, and upper branch irradiates  a micromachined mm-scale vapor cell. In each path, there are optical components that control the lasers' polarization (POL. CONT.), a photo-detector (PD), and a lock-in amplifier (LIA). }
\end{figure}
We describe our  experimental setup, depicted schematically in Figure ~\ref{fig:setup}, which consists of the two above mentioned  counter-propagating probe and pump beams,  each with a diameter of approximately 1 mm. Each beam is split into a measurement path and a reference path, which are nearly identical except for the vapor cells: one is a millimeter-scale micromachined cell, while the other is a standard glass-blown cell 70\,mm in length. In both paths the probe beam propagates through the cell and is then detected by a photodetector (PD). The blue beam enters from the opposite side through a dichroic mirror (DM) and counter-propagates through the cell, exciting the atoms to the Rydberg level. Each of the four beams pass through a half-wave plate and a neutral-density (ND) filter for polarization and intensity control, respectively. To improve noise screening, the blue laser's frequency is directly modulated at approximately 2 kHz. Our spectroscopic signal is achieved by demodulating the PD signals by means of lock-in amplifier (LIA) referenced to the modulation frequency.  In addition, we regulate the atomic density in the micromachined cell by adjusting its temperature using a resistive-heater based oven.  Our reference cell defines the zero-detuning point at the center of its Rydberg transition. In addition, we use the well known spectral distance between the two D-state Rydberg lines, namely  \(53D_{5/2}\) and  \(53D_{3/2}\), to calibrate the laser detuning frequency scale.

\section{Results}
A thorough characterization of the spectroscopic signal in the confined Pyrex–Si–Pyrex environment is critical for advancing noninvasive, field-deployable measurements. Accordingly, we examine the lineshape of the Rydberg transmission in the micromachined cell. Our measurements account for key system parameters, including temperature, optical intensity, and interrogation position. These measurements allow us to establish the sensor’s sensitivity limit, defined by the signal’s linewidth, contrast and noise, and also provide insight into the fundamental behavior of Rydberg atoms near surfaces in micromachined vapor cells. 

First, we perform Rydberg spectroscopy measurements at various temperatures. To control the microcell’s temperature, it is mounted on a resistive heater that regulates the temperature of the cell's baseplate, while the remaining surfaces are exposed to the ambient laboratory environment. This setup leads to substantial thermal gradients, which complicate accurate measurement of the cold spot temperature that governs the vapor pressure. To address this, we estimate the effective temperature by measuring absorption of the probe laser to infer the corresponding atomic density. This density is then correlated with temperature using a separately calibrated reference system comprising a closed-form oven and a thermistor.

\begin{figure}[ht]
\includegraphics[width=1\linewidth]{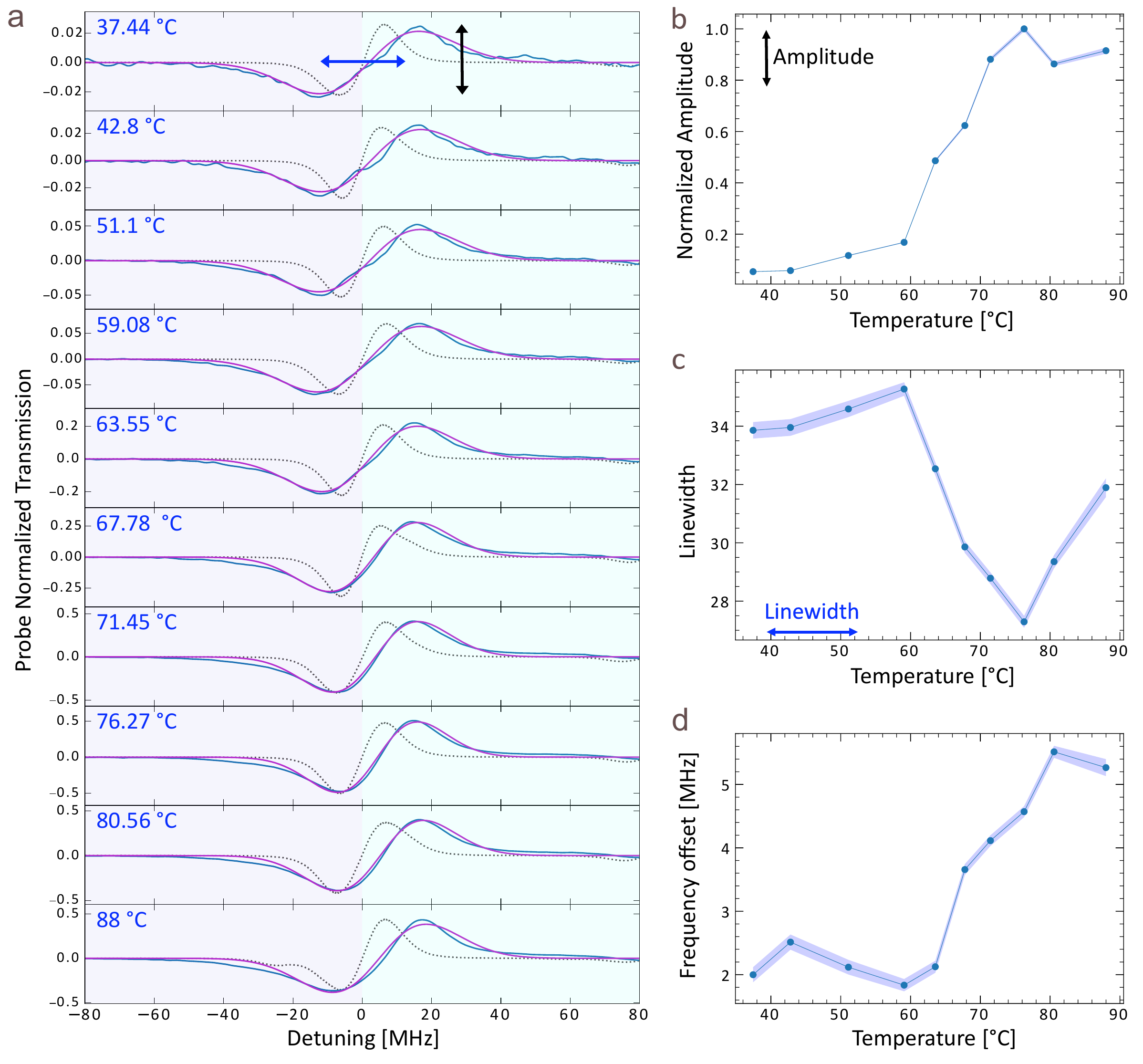}
\caption{\label{fig:temp}\textbf{Cell Temperature Dependence of Rydberg Spectra } (a) Probe transmission of a micromachined vapor cell at different atomic densities corresponding to different temperatures (dark blue). Also shown is the respective signal in the reference vapor cell (dotted grey). The signal is fitted to the derivative of a gaussian illustrated in purple. Sub-figures (b-c) represent the  amplitude (b), the linewidth (c) , and the offset of the signal compared to the reference cell (d) as a function of temperature. In sub-figures (b-d), the thickness of the line is determined by the confinement bounds of the fit.}
\end{figure}

Once the vapor cell is stabilized at the desired temperature, we proceed to model the spectral lines. We assume that the dominant broadening mechanism is Gaussian in nature, primarily due to residual Doppler effects, and therefore fit the derivative of a Gaussian function to the measured signal, consistent with the frequency-demodulated nature of our detection method. From these fits, we extract three key parameters: the linewidth (expressed as the full width at half maximum, FWHM), the peak-to-peak amplitude, and the spectral offset relative to a reference cell. The results are presented in Figure. ~\ref{fig:temp}. We note that the extracted linewidth corresponds to the original Gaussian profile, while the actual detection sensitivity is often determined by the slope of its derivative. As a result, the FWHM provides an upper bound, typically about 15 percent higher, on the effective slope relevant for field sensing.

As the temperature increases up to approximately 60°C, we observe a slight increase in the peak amplitude. Beyond 60°C and up to about 75°C, the amplitude rises steeply, followed by a subsequent decline and stabilization at higher temperatures (Figure~\ref{fig:temp} b). These temperature-dependent “turning points” are also reflected in the linewidth behavior. Initially, the line broadens slightly up to about 60°C, then narrows sharply between 60°C and 75°C before undergoing another sharp turning point at around 75°C and broadening steeply thereafter (Figure~\ref{fig:temp} c). In terms of the spectral offset, we fond little change up to roughly 65°C. Beyond this temperature, however, the line exhibits a pronounced shift, which continues until about 78°C (Figure~\ref{fig:temp} d).

We assume that these variations are related to DC Stark shifts caused by temperature-dependent electrostatic fields and spatial electric-field gradients within the cell. Thus, as the cell temperature changes, not only does the atomic density vary, but the internal DC fields within the cell are also altered. These changes could arise from temperature-dependent effects such as the Seebeck effect~\cite{lee2014seebeck, lee2013evaluating}, or other thermally induced phenomena like the migration of liquid Rb, within the cell. Indeed, liquid Rb may be charged  or even screen existing surface charge that resides within the cell walls. As discussed above, DC electric fields and spatial electric-field gradients can both shift and broaden the lineshape. Such effects stem from non-uniform Stark shift coefficients attributed to each \(m_j\) sub-level, and spatially varying electric fields contribute to shifting each lineshape differently; when these shifts are smaller than the linewidth, they primarily manifest as line broadening. We suggest that the electric field gradient minimizes around 75°C, reducing the linewidth, whereas the overall electric field increases, thus shifting the line center.

\begin{figure}[ht]
\includegraphics[width=1\linewidth]{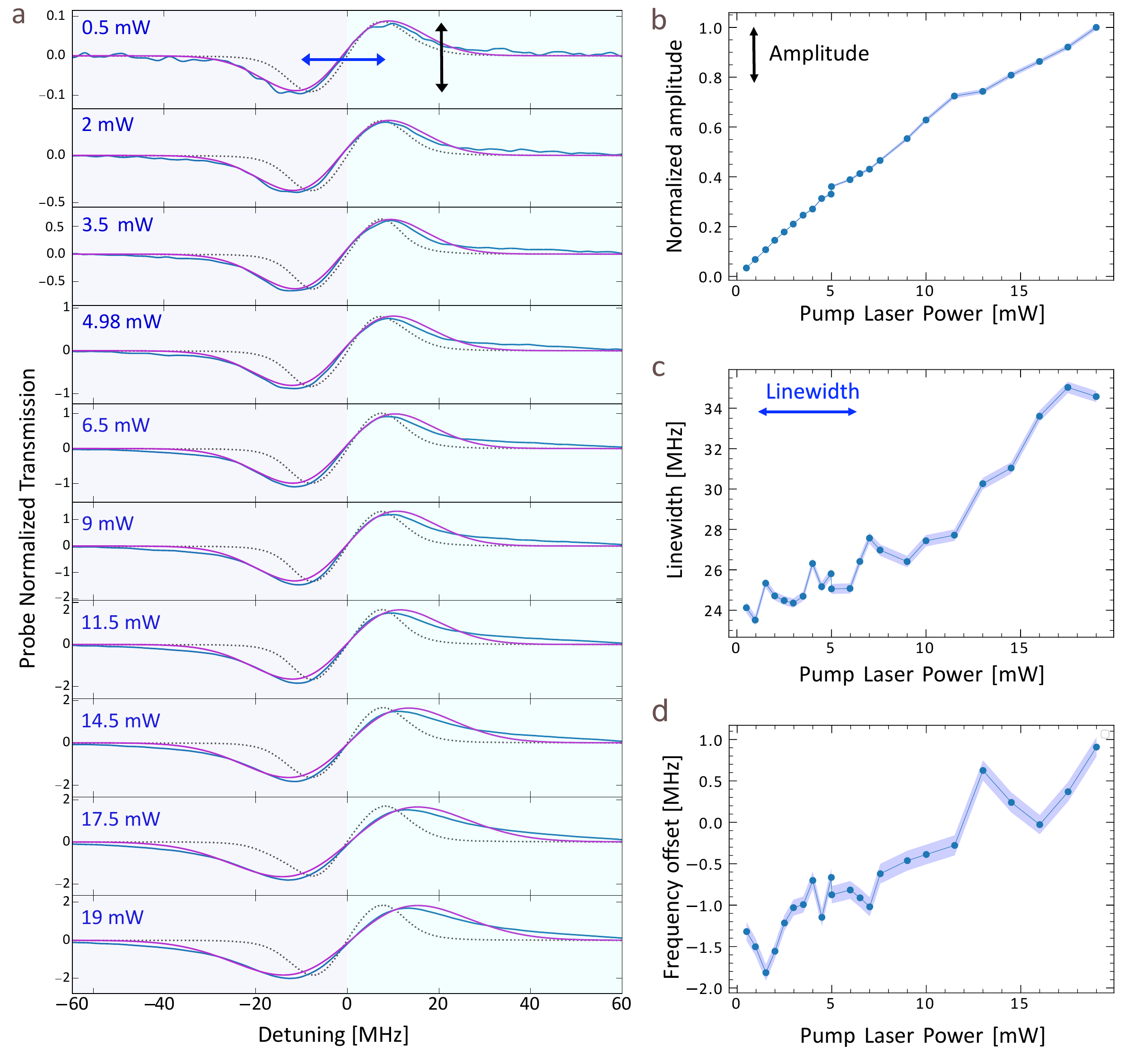}
\caption{\label{fig:power} \textbf{Impact of Pump Laser Power on Rydberg Spectra.} (a) Probe transmission through a micromachined vapor cell for various pump laser powers (dark blue), along with the corresponding signal from a reference vapor cell (dotted grey). The signal is fitted to the derivative of a Gaussian, shown in purple. Subfigures (b)–(d) display the amplitude (b), linewidth (c), and frequency offset relative to the reference cell (d) as functions of the pump laser power. In sub-figures (b-d), the thickness of the line is determined by the confinement bounds of the fit.}
\end{figure}

Next, we investigate how the signal changes as we vary the pump laser power. As before, we extract the linewidth, amplitude, and spectral offset from the measured signals (see FIg. \ref{fig:power}). Guided by our previous findings, we set the operating temperature to around 75°C, where the linewidth is relatively narrow and the signal-to-noise ratio is high. Under these conditions, we observe a steady increase in amplitude, linewidth, and frequency offset with rising pump power (Figures \ref{fig:power}(b), \ref{fig:power}(c), and \ref{fig:power}(d), respectively). We attribute the changes in linewidth and spectral shifts to the ionization of liquid Rb, leading to the creation of surface charge and subsequent DC Stark shifts and broadening. Similar results in cm-scale vapor cells have been reported in ref.  ~\cite{ma2020dc}. Our assumption is based on the fact that the work function of liquid rubidium is approximately \(2.1-2.3\ \mathrm{eV}\), meaning that the pump laser wavelength (\(480 \mathrm{nm}\)) provides sufficient energy to ionize rubidium residing on the cell’s inner surfaces.  This ionization can create a sheet of charge on the cell walls, altering the electrostatic field inside the microcell and thereby shifting and broadening the spectral line. Other possible mechanisms, such as light shifts or heating effects, seem unlikely  considering the transition dipole of the pump frequency. Indeed, this transition dipole element (and associated Rabi frequency) for our pump is relatively small, yielding light shifts significantly smaller than our measured shifts. Similarly, the fraction of blue light absorbed by the Pyrex windows is minimal and cannot produce the observed frequency shifts, aligning with our previously measured thermal behaviors (see Figure~\ref{fig:temp}).

We emphasize that the results presented above were obtained using a procedure where the cells are exposed to a thermal gradient, causing the liquid rubidium to migrate from the Pyrex windows toward the dispenser pill area. This migration results in relatively little liquid Rb being adsorbed onto the spectroscopy region. However, it is extremely informative to study the cell in a non-optimal state (i.e., prior to undergoing this thermal process). Such an approach allows us to further investigate the effects of liquid Rb on the Rydberg lineshapes and correlate these findings with our previous observations. Additionally, these measurements illustrate non-optimal charging effects and underscore the need to confine the liquid Rb -and the dispenser pill, as will be shown-to a designated area away from the spectroscopy region. This strategy has significant implications for other quantum sensors, such as TPA chip-scale clocks. 

\begin{figure}[ht]
\includegraphics[width=1\linewidth]{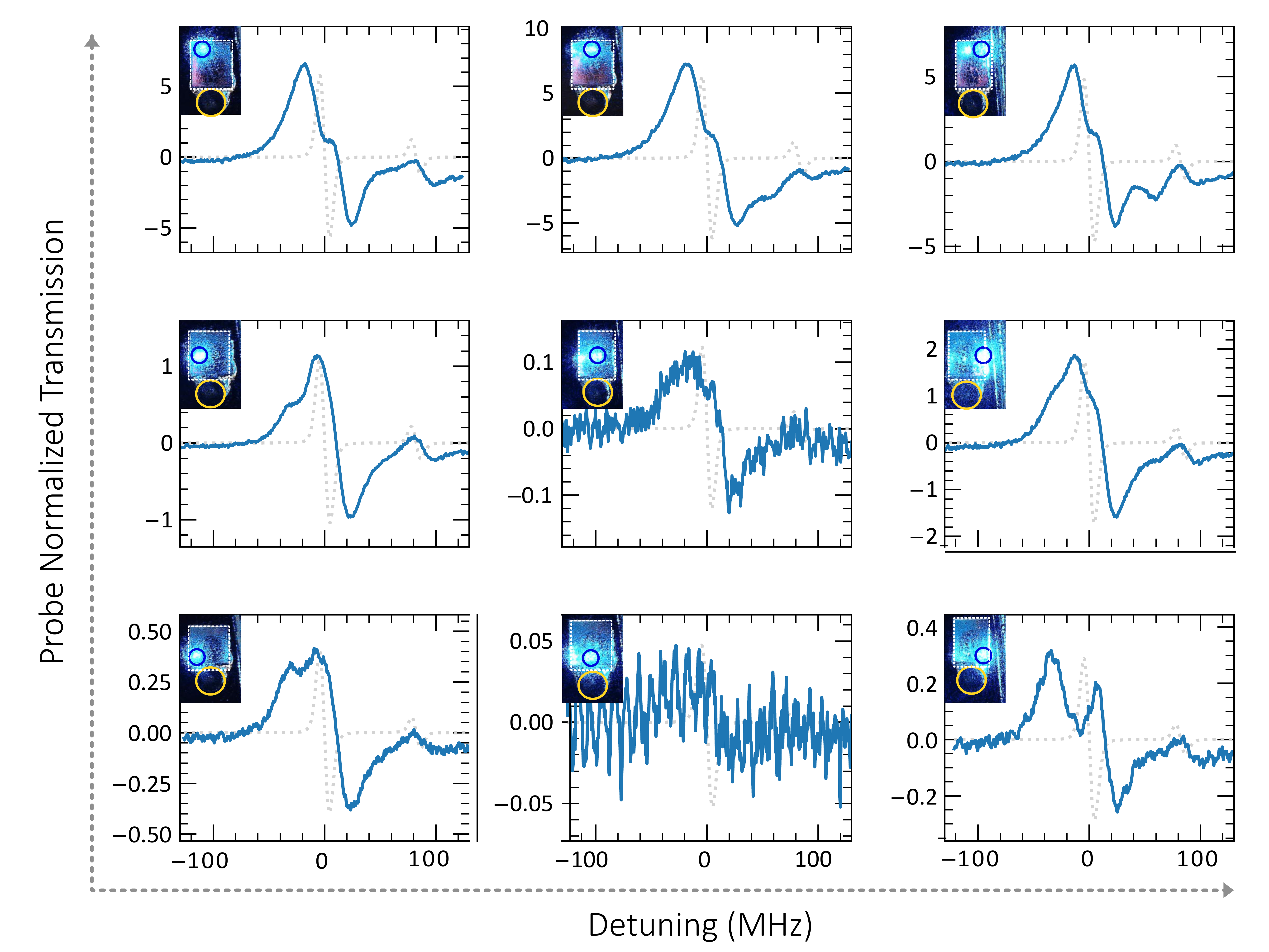}
\caption{\label{fig:spetial} \textbf{Mapping the Spatial Variation of Rydberg Spectra within the Cell.} Rydberg transmission lines acquired under non-optimal conditions (in terms of the Rb droplet position) at different areas of the micromachined vapor cell. The position of the laser beam relative to the cell area is presented for each signal; the beam location is circled in blue and the rubidium pill in yellow.}
\end{figure}

To investigate these non-optimal signals, we perform Rydberg spectroscopy while varying the interaction area of our laser excitation volume relative to the cell walls. Specifically, we systematically move the cell in small increments and record the Rydberg line at each position. We have mapped a 3×3 grid of cell positions, as shown in Figure ~\ref{fig:spetial}(a). For each measurement, the current position of the cell relative to the laser beams is displayed in the top left corner of each plot, with a blue circle marking the center of the beam and a yellow circle indicating the location of the Rubidium dispenser pill.

We observe strong variations in the spectroscopic signals depending on the position within the cell. In general, there are notable changes in linewidth, contrast, and splitting. These effects are attributed to DC Stark shifts, reinforcing our hypothesis of a spatial DC field profile inside the cell. Our cell features a Pyrex spectroscopy window through which the light passes, and a small etched circular region that holds the rubidium dispenser pill (indicated by the yellow circle) connected via a small aperture, as shown in Figure~\ref{fig:spetial} (b). We note that the spectral lineshape is most severely affected in regions of the cell that are closest to the rubidium pill; specifically, in the bottom center of the cell-where the distance to the pill is minimal-the signal becomes almost completely distorted.  It can therefore be speculated that the dispenser pill (which contains alloy of Zr and Al ~\cite{maurice2016design}) distorts the electrostatic environment around it, leading to significant broadening. Additionally, residues from the activation process residing near the dispenser pill as well as liquid Rb droplets may contribute to these electrostatic conditions and also create scattering centers that distort the optical wavefront, thereby reducing signal integrity.

\begin{figure}[h]
\includegraphics[width=1\linewidth]{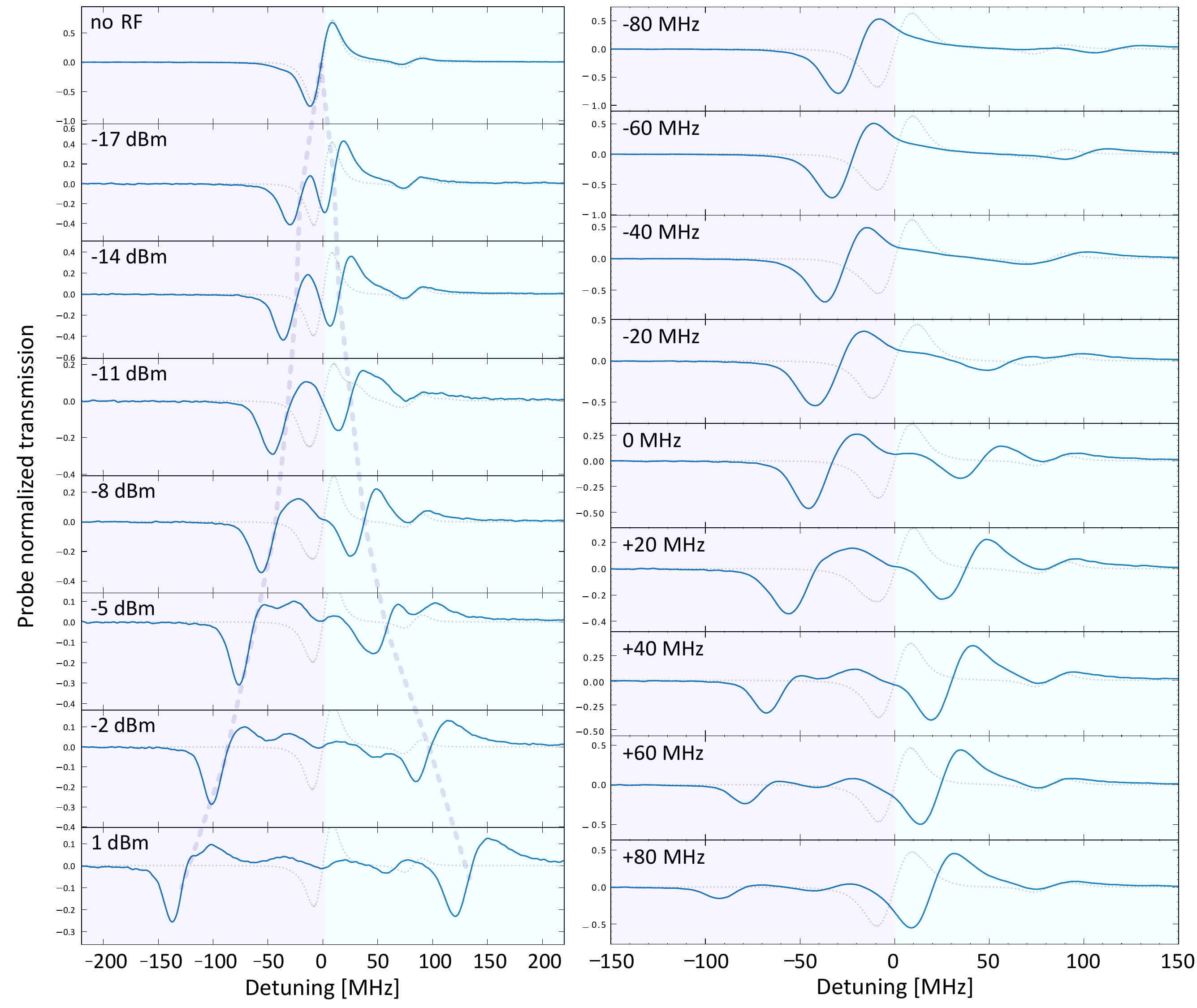}
\caption{\label{fig:RF} \textbf{RF-Induced Autler–Townes Splitting in a Micromachined Vapor Cell.} (a) AT splitting observed under different external RF fields. The coupling frequency is offset by 20 MHz from the theoretical value to achieve optimal symmetry. (b) AT splitting and light-shifts for different frequency offsets from the RF transition resonance under a fixed field power of -8 dBm; the most symmetrical signal is obtained at a 20 MHz offset.}
\end{figure}

After we have thoroughly characterized the Rydberg lineshape in our micromachined vapor cell, we turn to investigate RF electrometery in such cells. To that end, we couple the \(52D_{5/2}\) Rydberg state to the adjacent \(53P_{3/2}\) state using  a horn antenna excited by a synthesizer with frequency  around \(\approx 15.1 \ \mathrm{GHz}\). In Figure \ref{fig:RF}(a), we present the spectroscopic lineshapes of our micro-cell subject to varying electric-field powers represented by the reading of our synthesizer (i.e, no attempt has been to measure the actual coupling to the horn antenna). Clear AT splitting is observed, following a square-root dependence with respect to the RF  power launched to the antenna. However, we notice effects of asymmetry in the AT splittings which are strongly evident for the lower applied RF powers. For example, we can see in (Figure ~\ref{eq:ATsplitting} (a)) that for \(-17\ \mathrm{dBm}\) , the left linesahpe is visibly smaller than the right linshape. As the RF amplitude increases, and the splitting gets larger, we observe a gradual tendency to a symmetrical split, so that in strong fields such as \(1 \ \mathrm{dBm}\), we see a relatively symmetrical split. This asymmetry is likely a result of a small residual electric-field stemming from the aforementioned effects of remaining surface charge. This claim is reinforced, when considering that our results are obtained by offsetting the RF frequency from the calculated coupling frequency by approximately \(20 \ \mathrm{MHz}\).

We investigate further the effect of the RF frequency offset on the symmetry of the AT splittings. To do so, we fix the amplitude of the coupling RF field to a constant value, detune the RF frequency in small increments, and measure the AT splitting for each detuned RF frequency. These results are plotted in Figure ~\ref{fig:RF}(b) for AT splitting caused by a field of \(-8 \ \mathrm{dBm}\) at different offsets of the RF field, spanning from \(-80 \ \mathrm{MHz}\) to \(80 \ \mathrm{MHz}\). Clearly, the asymmetry is closely correlated with the the RF offset: in the case of no RF offset, we can easily detect the split, but it is evidently asymmetrical with a more prominent line at the left hand side. When the frequency is vary far from the the resonance, we transition to the regime of a pure AC-stark shift, often coined light-shift. Such behavior alludes again to the presence of an electrostatic potential shifting the energy gap between the \(52D_{5/2}\) Rydberg state to the adjacent \(53P_{3/2}\) state. Indeed, we calculate the weighted average of such frequency offset and find it to be consistent with electric-fields of  \(0.2 \ \mathrm{V/cm}\).

\FloatBarrier

\section{Discussion}

We have presented a chip-scale electric-field sensor based on a mm-scale micromachined Pyrex–Si–Pyrex vapor cell that enables all-optical, sub-wavelength RF sensing. Through careful examination and systematic characterization of key system parameters, we successfully achieved narrow-linewidth Rydberg spectroscopy in these ultra-compact vapor cells. The resulting spectroscopic response enables high-sensitivity RF detection with estimated sensitivities as low as \( 10\ \mu V/\text{cm}\). 

Our work focuses on identifying and addressing key challenges in transitioning from cm-scale to mm-scale Rydberg RF sensing. Our results indicate that spectral shifts, splitting, and broadening of the measured signal arise from the electrostatic environment within the cells, stemming from surface charge distributions. Additionally, spatially resolved measurements allowed us to identify distinct DC field profiles with areas near the dispenser pill exhibiting spectral shifts that can be attributed to electrostatic fields ranging from $0.2$ V/cm to $0.6$ V/cm. By optimizing the position of the interrogation area and managing the cell’s cold-spot temperature, we were able to reduce the linewidth and successfully demonstrate RF sensing via AT splitting. In this case, we estimate the electrostatic fields to be approximately $0.2$ V/cm, consistent with a spectral shift of the theoretical RF transition by $20$ MHz. Deviations from this coupling frequency result in asymmetrical spectral features, transitioning from pure AT splitting to light shifts.

We note that we were also able to demonstrate Rydberg spectroscopy using a heating laser at a wavelength of 450 nm, which is absorbed by the cell's silicon wall. In doing so, we achieved remote Rydberg sensing, enabling potential non-invasive, sub-wavelength spectroscopy with our micromachined vapor cells; this approach allows the cells to be positioned in a non-distorting environment (i.e., in the absence of metallic surfaces such as a resistive heater).
To estimate the sensitivity of our Rydberg sensor, we calculated the standard deviation of our off-resonant probe light, which represents the measured noise over a bandwidth corresponding to approximately 100 s. We note that these relatively long time scales are not a fundamental property of our system, as we did not optimize parameters such as light collection, temperature stability, or frequency stability. We chose this duration to achieve consistent results across the various parameter spaces we explored. By calculating the slope of the atomic response and multiplying its reciprocal by the aforementioned noise, we obtain a noise level corresponding to a sensitivity as low as \(10 \mu\)V/cm, representing a lower bound on the detectable field strength, and thus an upper bound on the sensor’s sensitivity.

Identification of the electrostatic spatial profile within micromachined vapor cells has important implications for other types of micromachined based quantum sensors. For instance, recent emerging chip-scale two-photon-absorption based (TPA) atomic clocks have already shown frequency instability floors of \(4\cdot10^{-15}\) at 1000 s integration time~\cite{newman2021high}. Yet, Martin et. al have estimated a DC-stark shift sensitivity of \(5.5\cdot 10^{-15}(V/cm)^2\) of such a TPA signal, thus accounting for $0.6 \ \mathrm{V/cm}$ electric field variations potentially resulting in frequency shifts of the order of \(2\cdot10^{-15}\), thus, limiting the stability and accuracy floor of such a clock~\cite{martin2019frequency}.

Our micromachined vapor cells consist of a Pyrex–Si–Pyrex anodically bonded structure. Pyrex has a relatively low permittivity compared to silicon’s much higher permittivity - a difference that influences the scattering properties of RF fields and distorts the RF field distribution at the detection volume through effects such as standing waves, reflection, and polarization distortion ~\cite {noaman2023vapor}. Moreover, doped Si may attenuate RF signals as they propagate through the Si surfaces. In addition, the adsorption of Rb on each surface can differ - Pyrex is known to have a higher tendency for adsorption relative to silicon, which can be detrimental for electrometry by broadening and shifting the lineshapes. To that end, future investigations of alternative material compositions such as sapphire-based cells, which are known to exhibit lower adsorption relative to Pyrex, could be beneficial. Currently, our cells have a Pyrex-to-silicon volume ratio of approximately three; however, increasing this ratio or using lower-doped silicon may further enhance signal integrity and reduce RF field distortion.

In summary, we have studied the spectroscopic properties of Rydberg atoms within micromachined mm-scale vapor cells. Our sub-wavelength interaction volume - an order of magnitude smaller than the RF wavelength, enables sensitive and non-invasive electrometry. A key aspect of our study is the thorough investigation of the role of surface charges in the spectroscopic lineshape, as well as in the underlying physics and mitigation strategies. Our results highlight the potential of micromachined vapor cells for precise electromagnetic field measurements with applications in communications, near-field RF imaging, and emerging chip-scale quantum technologies.

\begingroup
\small

\section*{Disclosures}
The authors declare no conflicts of interest.

\section*{Data Availability}
The data that support the findings of this study are available from the corresponding author upon reasonable request.\endgroup

\FloatBarrier
\newpage
\bibliography{references}
\end{document}